\documentclass[12pt]{article}

\begin{document}

\Large
\begin{center}

   \vspace*{3ex}
     {\bf  Super quantum mechanics, spacetime and quantum information  } \\
             \vspace*{2ex}
              {\bf Jaroslav Hruby}

   {\bf Institute of Physics AV CR,Czech Republic }
  {\normalsize \it e-mail: hruby@fzu.cz  }


\vspace*{2ex}
\begin{abstract}

    In the paper it is shown  the connection between spacetime
and quantum information using an information spacetime as
superspace. The connection of quantum information with
anticommuting variables is given . Also a solitonic bag is
presented where the information is confined.

\end{abstract}

\end{center}

\vspace*{3ex}

\normalsize
\section{Some ideas from quantum information and super quantum mechanics }

       A quantum bit (qubit) is a quantum system with a two-dimensional
Hilbert space.

More precisely a qubit is the amount of the information which is
contained in a pure quantum state from the two-dimensional Hilbert
space ${\cal H}_2$.

A general superposition state of the qubit is
\begin{equation}
 |\psi\rangle  = \psi_0 |0\rangle  + \psi_1 |1\rangle  ,
\label{3.1}
\end{equation}
where  $\psi_0$  and   $\psi_1$ are complex numbers, $|0\rangle$
and $|1\rangle$ are kets representing two Boolean states. The
superposition state has the propensity to be a $0$ or a $1$ and
${|\psi_0|}^2 + {|\psi_1|}^2 = 1$.

The eq.(1) can be written as
\begin{equation}
 |\psi\rangle  = {\psi}_0\left( \begin{array}{c} 1 \\ 0 \end{array} \right)
 + {\psi}_1 \left( \begin{array}{c} 0 \\ 1 \end{array} \right)  ,
\label{3.2}
\end{equation}

where we labelled $\left( \begin{array}{c} 1 \\ 0 \end{array}
\right) $ and $\left( \begin{array}{c} 0 \\ 1 \end{array} \right)$
two basis states zero and one.

The Clifford algebra relations of the $2\times2$ Dirac matrices is
\begin{equation}
\left\{{\gamma}^\mu,{\gamma}^\nu \right\} = 2 {\eta}^{\mu\nu} ,
\label{3.3}
\end{equation}
where
\begin{equation}
 {\eta}^{\mu\nu}    =  \left( \begin{array}{cc}
                              -1 & 0 \\
                               0 & 1 \end{array} \right).
\label{3.4}
\end{equation}

We choose the representation
\begin{equation}
 {\gamma}^0 =i{\sigma}^2   =  \left( \begin{array}{cc}
                              0 & 1 \\
                             -1 & 0 \end{array} \right)
\label{3.5}
\end{equation}
and
\begin{equation}
 {\gamma}^1 ={\sigma}^3   =  \left( \begin{array}{cc}
                              1 & 0 \\
                              0 &-1  \end{array} \right) ,
\label{3.6}
\end{equation}
where $\sigma$ are Pauli matrices and
${\gamma}^5={\gamma}^0{\gamma}^1$ .

We shall now assume, that quantum information is connected with an
anticommuting variable.So we obtain from the super quantum
mechanics.

The glue between quantum mechanics and supersymmetry was first
given by the supersoliton Lagrangian in $(1+1)$ space-time
dimensions \cite{1} in~1977
\begin{equation}
 L =
\frac{1}{2}[(\partial_{\mu}\phi)^{2}-V^{2}(\phi)+
        \bar{\psi}(i+V'(\phi))\psi]    \label{v1}
\end{equation}
where $\phi$ was a solitonic Bose field and $\psi$ was a Fermi
field and $V(\phi)$ was some nonlinear potential. Supersymmetric
quantum mechanics (super quantum mechanics SQM) follows directly
via reduction from $(1+1)$ space-time dimensions to $(0+1)$
dimension \cite{2}.

SQM was established by E.Witten \cite{3} and P. Salamonson a J. W.
Van Holten  \cite{4} and corresponding Hamiltonian SSQM has the
form:
\begin{equation}
 H = \frac{1}{2} p^{2}+\frac{1}{2}V^{2}(x)
   -\frac{[\psi^{*},\psi]}{2}V'(x)                 \label{1}
\end{equation}
where \([p,x]=-i\), \(\{\psi^{*},\psi\}=1\) and $V(x)$ is a
potencial, where $p$, $x$ are Bose operators of impuls and
space-coordinate and $\psi$, $\psi^{*}$ are Fermi variables.

Following the idea that space-time and information are connected
as a fiber space we shall show \cite{5} how it will be work for
SQM.

We shall use superfield formalism on the superspace
$(t,\theta,\theta^{*})$, where Bose variable is the time and
$\theta$, $\theta^{*}$ are Grassmannian anticommuting variable:
\begin{equation}
  \{\theta,\theta\} = \{\theta^{*},\theta^{*}\} = 0 \ , \qquad
  [\theta,t]=0 \ , \qquad \{\theta,\theta^{*}\} = 1 \ .
                                                     \label{2}
\end{equation}

Supersymmetric transformation has the form:
\begin{eqnarray}
 t' & = & t-i(\theta^{*}\varepsilon-\varepsilon^{*}\theta) \ , \nonumber \\
 \theta' & = & \theta+\varepsilon         \ , \label{3dupl} \\
 \theta^{*'} & = & \theta^{*}+\varepsilon^{*}     \ , \nonumber
\end{eqnarray}
and the superalgebra generators are:
\begin{eqnarray}
 Q & = & i\partial_{\theta}-\theta^{*}\partial_{t}    \ , \label{3} \\
 Q^{*} & = & -i\partial_{\theta^{*}}+\theta\partial_{t}     \nonumber
\end{eqnarray}
We can see that
\begin{eqnarray}
 \{Q,Q^{*}\} & = & 2i\partial_{t}=2H   \ , \label{4} \\
 \{Q,H\} & = & 0                     \nonumber
\end{eqnarray}

Supercovariant derivatives have the form
\begin{equation}
  D_{\theta} = \partial_{\theta}-i\theta^{*}\partial_{t} \ ,
  \qquad   D_{\theta^{*}} = \partial_{\theta^{*}}-i\theta\partial_{t} \ .
                                                     \label{5}
\end{equation}

We shall define a scalar superfield:
\[ \Phi(t,\theta,\theta^{*})=\Phi^{*}(t,\theta,\theta^{*}) \ , \]
and the expansion in $\theta$, $\theta^{*}$ is:
\begin{equation}
 \Phi(t,\theta,\theta^{*})=x(t)+i\theta\psi(t)-i\psi^{*}(t)\theta^{*}
   +\theta^{*}\theta D(t) \ .       \label{6}
\end{equation}

The coefficients in the expansion (\ref{6}) are in SQM Bose
one-dimensional space variable $x(t)$ and Fermi variable $\psi(t)$
connected with the qubit information. The component $D$ can be
interpreted as a potencial, as we shall see.

The variation
\begin{eqnarray}
 \delta\Phi & = & i[\varepsilon^{*}Q^{*}+Q\varepsilon,\Phi]  \nonumber \\
   & = & \delta x(t)+i\theta\delta\psi(t)
   -i\delta\psi^{*}(t)\theta^{*}+
    \theta^{*}\theta\delta D(t)              \label{7}
\end{eqnarray}
a gives us the supersymmetric transformations as follows:
\begin{eqnarray*}
 i\delta x & = & \varepsilon^{*}\psi^{*}-\psi\varepsilon \ ,       \\
 \delta\psi & = & -i\varepsilon^{*}D+\varepsilon^{*}\dot{x} \ ,       \\
 \delta D & = & \frac{\partial}{\partial t}(\varepsilon\psi+
    \psi^{*}\varepsilon^{*}) \ .
\end{eqnarray*}

We shall assume that a function $f(\Phi)$ can be expanded as:
\begin{equation}
 f(\Phi)=\sum_{n} a_{n}\Phi^{n}               \label{8}
\end{equation}
and the action has the form:
\begin{equation}
  S=\int dt\,d\theta^{*}\,d\theta\,\left(
  \frac{1}{2}(D_{\theta}\Phi)^{2}-f(\Phi)\right) \ ,   \label{9}
\end{equation}
because $dt\,d\theta^{*}\,d\theta$ is invariant and we assume only
the second order derivatives of the superfield components.

We shall use  Berezin´s integral:
\[ \int \theta\,d\theta= \int \theta^{*}\,d\theta^{*}=1 \ , \qquad
  \int d\theta= \int d\theta^{*}=0 \ ,         \]
and for the construction of Lagrangian we look for coefficient
with $\theta\theta^{*}$.

From the product:
\[ (D_{\theta}\Phi)^{*}D_{\theta}\Phi=
 (-i\psi^{*}-\theta D+i\theta\dot{x}+\theta^{*}\theta\dot{\psi}^{*})\times
 (i\psi-\theta^{*}D-i\theta^{*}\dot{x}+\theta\theta^{*}\dot{\psi}) \]
we can see that $\theta\theta^{*}$ coefficient is
\[ (\dot{x}^{2}+i(\psi^{*}\dot{\psi}-\dot{\psi}^{*}\psi)+D^{2}) \ .   \]

For the coefficient with $\theta\theta^{*}$ in the expansion of
$f(\Phi)$ we get:
\begin{eqnarray*}
 \lefteqn{ \biggl\{ \sum na_{n}x^{n-1}(-D)+\sum n(n-1)a_{n}x^{n-2}
 \left[ \frac{1}{2}(\psi\psi^{*}-\psi^{*}\psi) \right] \biggr\}=}  \\
   & = & \left\{ -Df'(x)-\frac{1}{2}[\psi^{*},\psi]f''\right\} \ .
\end{eqnarray*}

After integration in the action(\ref{10}) in variables
$\theta\theta^{*}$, we get:
\begin{equation}
  S=\int dt\,\left(\frac{1}{2}\dot{x}^{2}+
  \frac{1}{2}i(\psi^{*}\dot{\psi}-\dot{\psi}^{*}\psi)
  +\frac{1}{2}D^{2}+Df'(x)+\frac{[\psi^{*},\psi]}{2}f''\right) \ . \label{11}
\end{equation}

We can eliminate the component \(D=-f'(x)=V(x)\), as is usual and
so we get the SQM Hamiltonian in the form (\ref{1}).

\section{Supersoliton as a bag for quantum information }

In the work \cite{5} we show a model of spacetime and information
like a fiber space. Following this idea in every moment in super
time $(t,\theta,\theta^{*})$, we can define a supersoliton field
$S(x,\theta)$, which is the solution for example of the super
sine-Gordon equation:
\[ \frac{i}{2}\,\bar{D}D\,S(x,\theta)=-\frac{a}{b}\sin{b}S(x,\theta) \ , \]
where we take the constants \(a=2i\sqrt{\alpha_{0}}, \ b=\frac{\beta}{2}\) , \\
as is usual in sine-Gordon model. The supersoliton has the form:
\begin{equation}
 S(x,\theta) = \frac{4}{\beta}\,\mbox{tg}^{-1}\exp{\left[
  \sqrt{\alpha_{0}}\,x+i
   \frac{\beta}{2\sin{\frac{\beta}{2}}\varphi_{\mbox{\tiny cl.}}(x)}
   \,\bar{\theta}\psi_{\mbox{\scriptsize cl.}}(x)-
   \frac{\sqrt{\alpha_{0}}}{2}\bar{\theta}\theta
   \right]} \ .                            \label{4.6}
\end{equation}

We can see
\begin{eqnarray}
 S(x,\theta) & = & \frac{4}{\beta}\,\mbox{tg}^{-1}\left[
 e^{\sqrt{\alpha_{0}}\,x}\left(1+
   i\frac{\beta}{2\sin{\frac{\beta}{2}}\varphi_{\mbox{\tiny cl.}}(x)}
   \,\bar{\theta}\psi_{\mbox{\scriptsize cl.}}(x)-
   \frac{\sqrt{\alpha_{0}}}{2}\bar{\theta}\theta
   \right)\right] \              \nonumber \\
   & = & \frac{4}{\beta}\,\mbox{tg}^{-1}e^{\sqrt{\alpha_{0}}\,x}
   +\frac{4}{\beta}\frac{e^{\sqrt{\alpha_{0}}\,x}}
   {1+e^{2\sqrt{\alpha_{0}}\,x}}\left(
   \frac{i\beta\,\bar{\theta}\psi_{\mbox{\scriptsize cl.}}(x)}
    {2\sin{\frac{\beta}{2}}\varphi_{\mbox{\scriptsize cl.}}(x)}
   -\frac{\sqrt{\alpha_{0}}}{2}\bar{\theta}\theta
   \right) \ .                   \label{4.7}
\end{eqnarray}
Because:
\(\frac{e^{\sqrt{\alpha_{0}}\,x}}{1+e^{2\sqrt{\alpha_{0}}\,x}}=
   \frac{\mbox{tg}\,\frac{\beta}{4}\,\varphi_{\mbox{\tiny cl.}}(x)}
    {1+\mbox{tg}^{2}\,\frac{\beta}{4}\,\varphi_{\mbox{\tiny cl.}}(x)}=
 \frac{1}{2}\,\sin{\frac{\beta}{2}}\,\varphi_{\mbox{\scriptsize cl.}}(x)
   \ , \) \\
we get:
\begin{equation}
 S(x,\theta) = \varphi_{\mbox{\scriptsize cl.}}(x)
  +i\bar{\theta}\psi_{\mbox{\scriptsize cl.}}(x)
  +\frac{\sqrt{\alpha_{0}}}{2}\,\bar{\theta}\theta\,F(x) \ ,
                                        \label{4.8}
\end{equation}
where F:
\[ F=\frac{2i\sqrt{\alpha_{0}}}{\beta}\sin{\frac{\beta}{2}}
 \varphi_{\mbox{\scriptsize cl.}} \ . \]

\medskip
Supersoliton solution(\ref{19}) can be obtained as follows:
we use
\(e^{y}=\sum_{k=0}^{\infty}\frac{y^{k}}{k!}\),  \\
where \(y=\left[\sqrt{\alpha_{0}}\,x
 +i\,\frac{\beta}{2\sin\frac{\beta}{2}
  \varphi_{\mbox{\tiny cl.}}}\,\bar{\theta}\psi_{\mbox{\scriptsize cl.}}
  -\frac{\sqrt{\alpha_{0}}}{2}\bar{\theta}\theta \right]\).

Because $\theta$ is anticommuting and we know that
\(\bar{\psi}_{\mbox{\scriptsize cl.}}\psi_{\mbox{\scriptsize
cl.}}=0\), we get only $e^{\sqrt{\alpha_{0}}\,x}$, and all other
are zero ( we use \(\bar{\theta}\psi=\bar{\psi}\theta\)) and so we
obtain:
\[ e^{\sqrt{\alpha_{0}}\,x} \left(
   1+i\,\frac{\beta}{2\sin\frac{\beta}{2}
  \varphi_{\mbox{\tiny cl.}}}\,\bar{\theta}\psi_{\mbox{\scriptsize cl.}}
  -\frac{\sqrt{\alpha_{0}}}{2}\bar{\theta}\theta
     \right) = \mbox{def.}\ z \ , \]
where def. $z$ means definition of the $z$.

Now we expand the \(f(z)=\mbox{arctg}\,z\)
\begin{eqnarray*}
 f(z) & = & f(\sigma)+f'(\sigma)(z-\sigma)+f''(\sigma)(z-\sigma)^{2}+\ldots
  \ ,         \\
  o & = & e^{\sqrt{\alpha_{0}}\,x} \ ,     \\
  (z-\sigma) & = & e^{\sqrt{\alpha_{0}}\,x}\left(
  i\,\frac{\beta}{2\sin\frac{\beta}{2}
  \varphi_{\mbox{\scriptsize cl.}}}\,\bar{\theta}\psi_{\mbox{\scriptsize cl.}}
  -\frac{\sqrt{\alpha_{0}}}{2}\bar{\theta}\theta \right) \ .
\end{eqnarray*}

We can see that from $(z-\sigma)^{2}$ all is equal zero.

Because \(f'(\sigma)=\frac{1}{1+\sigma^{2}}\), we get
\[  f(z)=f(e^{\sqrt{\alpha_{0}}\,x})
 +\frac{e^{\sqrt{\alpha_{0}}\,x}}{1+e^{2\sqrt{\alpha_{0}}\,x}}
  \left(i\,\frac{\beta}{2\sin\frac{\beta}{2}
  \varphi_{\mbox{\tiny cl.}}}\,\bar{\theta}\psi_{\mbox{\scriptsize cl.}}
  -\frac{\sqrt{\alpha_{0}}}{2}\bar{\theta}\theta  \right) \ .  \]

We can see
\[ \varphi_{\mbox{\scriptsize cl.}}= \frac{4}{\beta}\,\mbox{arctg}\,
 e^{\sqrt{\alpha_{0}}\,x} \ , \]
 and so \(e^{\sqrt{\alpha_{0}}\,x}=\mbox{tg}\,\frac{\beta}{4}
  \varphi_{\mbox{\scriptsize cl.}}\).

So we can write:
\[\frac{e^{\sqrt{\alpha_{0}}\,x}}{1+e^{2\sqrt{\alpha_{0}}\,x}}=
   \frac{\mbox{tg}\,\frac{\beta}{4}\,\varphi_{\mbox{\tiny cl.}}(x)}
    {1+\mbox{tg}^{2}\,\frac{\beta}{4}\,\varphi_{\mbox{\tiny cl.}}(x)}=
 \frac{1}{2}\,\sin{\frac{\beta}{2}}\,\varphi_{\mbox{\scriptsize cl.}}(x)
   \ .                 \]

for the solution $\psi_{\mbox{\scriptsize cl.}}(x)$ we can see
\[ \bar{\psi}_{\mbox{\scriptsize cl.}}\bar{\psi}_{\mbox{\scriptsize cl.}}=
  \psi_{\mbox{\scriptsize cl.}}^{+}\gamma^{0}\psi_{\mbox{\scriptsize cl.}}
  = C^{2}(\cosh{\sqrt{\alpha_{0}}\,x})^{-2}(1,i)
  \left(\begin{array}{rl}
   0 & 1 \\ 1 & 0
 \end{array}\right)
  {1\choose -i}=0          \]
and also
\[ \bar{\psi}_{\mbox{\scriptsize cl.}}\gamma^{1}
  \frac{d}{dx}\psi_{\mbox{\scriptsize cl.}}=
  \psi_{\mbox{\scriptsize cl.}}^{+}\gamma^{0}\psi_{\mbox{\scriptsize cl.}}
  = C^{2}(\cosh{\sqrt{\alpha_{0}}\,x})^{-2}(1,i)
  \left(\begin{array}{rl}
   -1 & 0 \\ 0 & 1
 \end{array}\right)
  {1\choose -i}=0 \ .         \]

It means $\psi_{\mbox{\scriptsize cl.}}(x)$ play no role for
energy:
\[ E_{\mbox{\scriptsize cl.}}(\varphi_{\mbox{\scriptsize cl.}},
  \psi_{\mbox{\scriptsize cl.}})=
  E_{\mbox{\scriptsize cl.}}(\psi_{\mbox{\scriptsize cl.}}) \ . \]

It is something like information field $\psi_{\mbox{\scriptsize
cl.}}(x)$ is in in a bag \cite{1}.
 \pagebreak

\section{Conclusions}

 Here we show another aspects of the connection of the spacetime
 and quantum information. We show the nontrivial connection of quantum information and
 spacetime via super quantum mechanics and the supersoliton theory,
 starting from the idea of the connection information and space
 time like superspace as a fiber space.
 It is interesting that such anticommuting information variables
 play no role in energy of the bag, where are confined.

   This work was supported by Grant T300100403 GA AV CR  .

\end{document}